\documentclass[a4paper,11pt]{article}
\usepackage{pos}

\title{Pionic depth of the hadron gas after a heavy-ion collision}

\author[a]{
Guillermo G\'omez Fonfr\'{\i}a, Felipe J. Llanes-Estrada, Javier Su\'arez Sucunza }
\author[b]{Juan M. Torres-Rincon}

\affiliation[a]{Univ. Complutense de Madrid, Dept. F\'{\i}sica Te\'orica, Plaza de las Ciencias 1, 28040 Madrid, Spain}

\affiliation[b]{Goethe Univ. Frankfurt, Inst. für Theoretische Physik,  Max-von-Laue-Str. 1, 60438 Frankfurt, Germany}

\emailAdd{fllanes@fis.ucm.es}
\emailAdd{torres-rincon@itp.uni-frankfurt.de}

\abstract{The final stage of a relativistic heavy-ion collision is a hadron gas. Final-state interactions therein distort the $p_T$ spectrum of particles coming from the phase transition upon cooling the quark-gluon plasma.
Using recent state-of-the-art parametrizations of pion interactions we provide theoretical computations of the pionic depth of the gas: how likely is it that a given pion rescatters in it (we find a high probability around $p_T=0.5$ GeV at midrapidity, corresponding to the formation of the $\rho$ resonance),  a comparison of the collision and  Bjorken expansion rates, and how many pions make it through without interacting as a function of $p_T$. This is in the range 10-24\% and shown in this plot, the main result of the contribution. 
\centerline{\includegraphics[width=0.55\columnwidth]{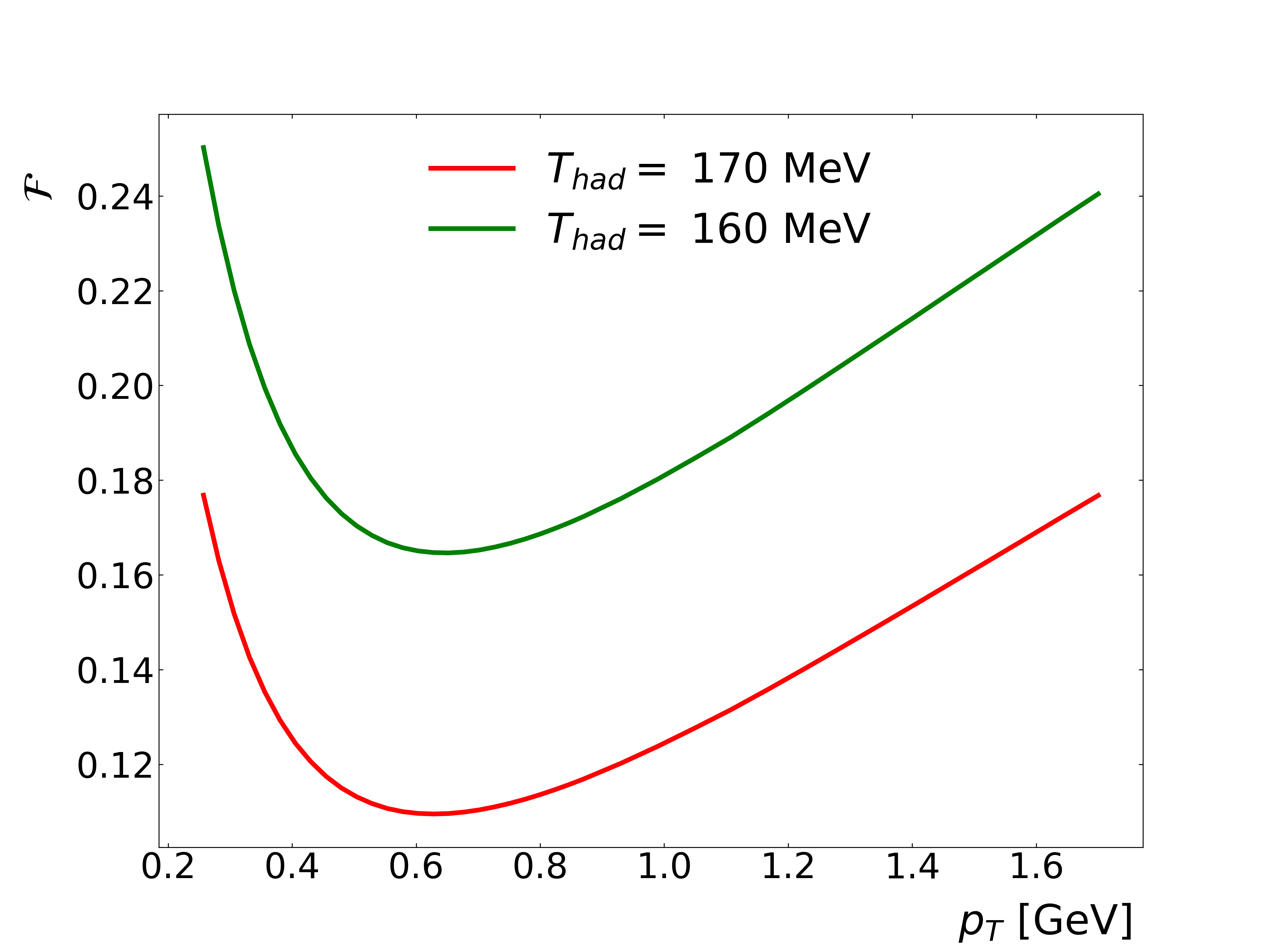}}}

\FullConference{%
  *** The European Physical Society Conference on High Energy Physics (EPS-HEP2021), ***\\
  *** 26-30 July 2021 ***\\
  *** Online conference, jointly organized by Universität Hamburg and the research center DESY ***
}

\begin{document}
\maketitle

It is of interest to ascertain the depth of the pion gas formed in the final state of heavy-ion collisions. This coldest phase, while the strong matter is decoupling, distorts the final-state spectra that are measured in detectors and veils the earlier stages of the collisions. Penetrating probes such as photons, dileptons, charmonia or jets are routinely used to more sharply view the quark-gluon plasma, but there is also abundant data with softer, more strongly-interacting particles that can be used. The effect of the late stage phase must be well understood to eventually apply corrections.

\begin{figure}[h]
\includegraphics[width=0.49\textwidth]{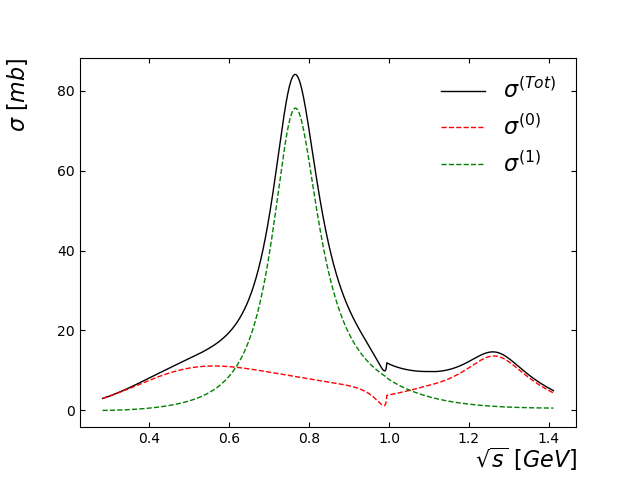}
\includegraphics[width=0.47\textwidth]{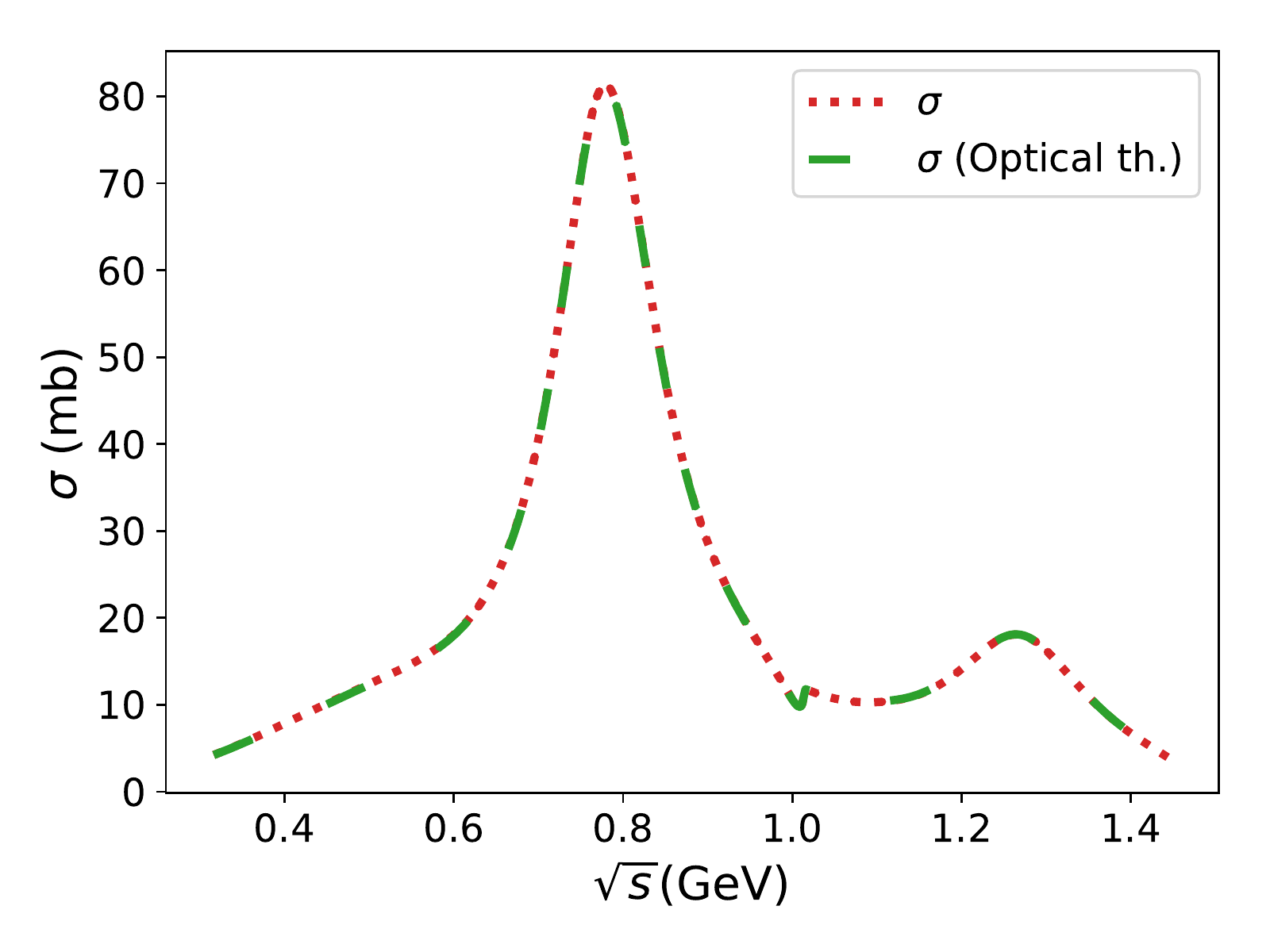}
\caption{\label{fig:sigma}
{\bf Left panel:} Breakdown of the $\pi\pi$ cross-section into definite isospin ($I=0,1$) contributions, $\sigma^{(I)}$ (we have also added $I=2$ that changes very little). {\bf Right panel:} Graphical check of the optical theorem, computing the $\sigma^{\textrm{(Tot)}}$ first by integrating over the solid angle and alternatively, from the forward amplitude. Various low-energy elastic resonances ($\sigma$, $\rho$, $f_0(980)$ and $f_2(1270)$ are clearly visible.)
}
\end{figure}

Fig.~\ref{fig:sigma} shows the $\pi\pi$ cross section through the relevant energies, broken down by isospin channels parametrized following~\cite{Kaminski:2012bv}, each of whose partial-waves separately satisfy elastic unitarity,
\begin{equation}
\sigma(s)= \frac{32\pi}{s} \sum_J (2J+1) \arrowvert t_{IJ} (s) \arrowvert^2 \ , \qquad
\mathrm{Im }[t_{IJ} (s)]  = \sqrt{1 - \frac{4 M_\pi^2}{s}} |t_{IJ}(s)|^2\ .
\end{equation}

\begin{figure}[h]
\includegraphics[width=0.48\columnwidth]{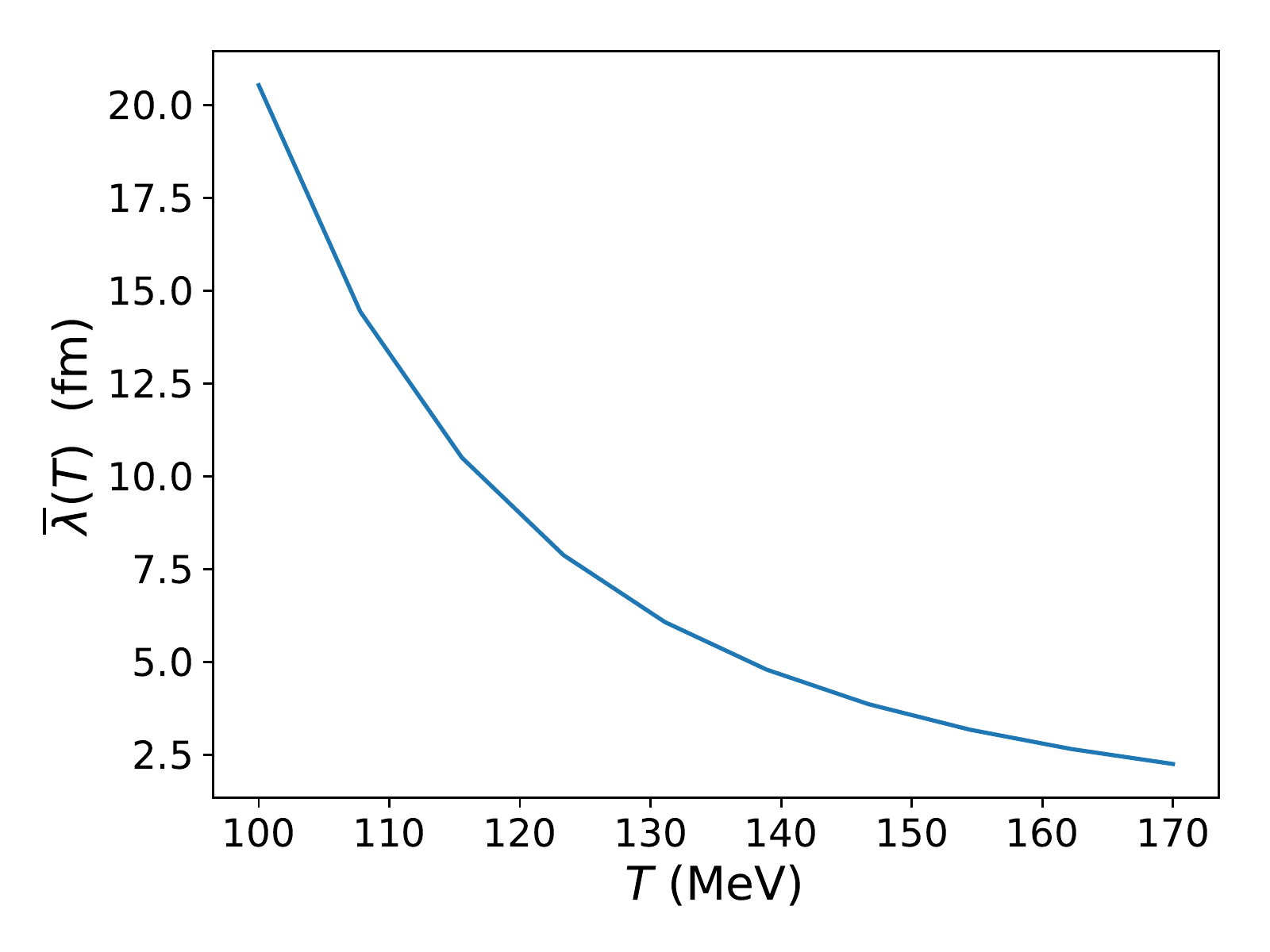}
\includegraphics[width=0.48\columnwidth]{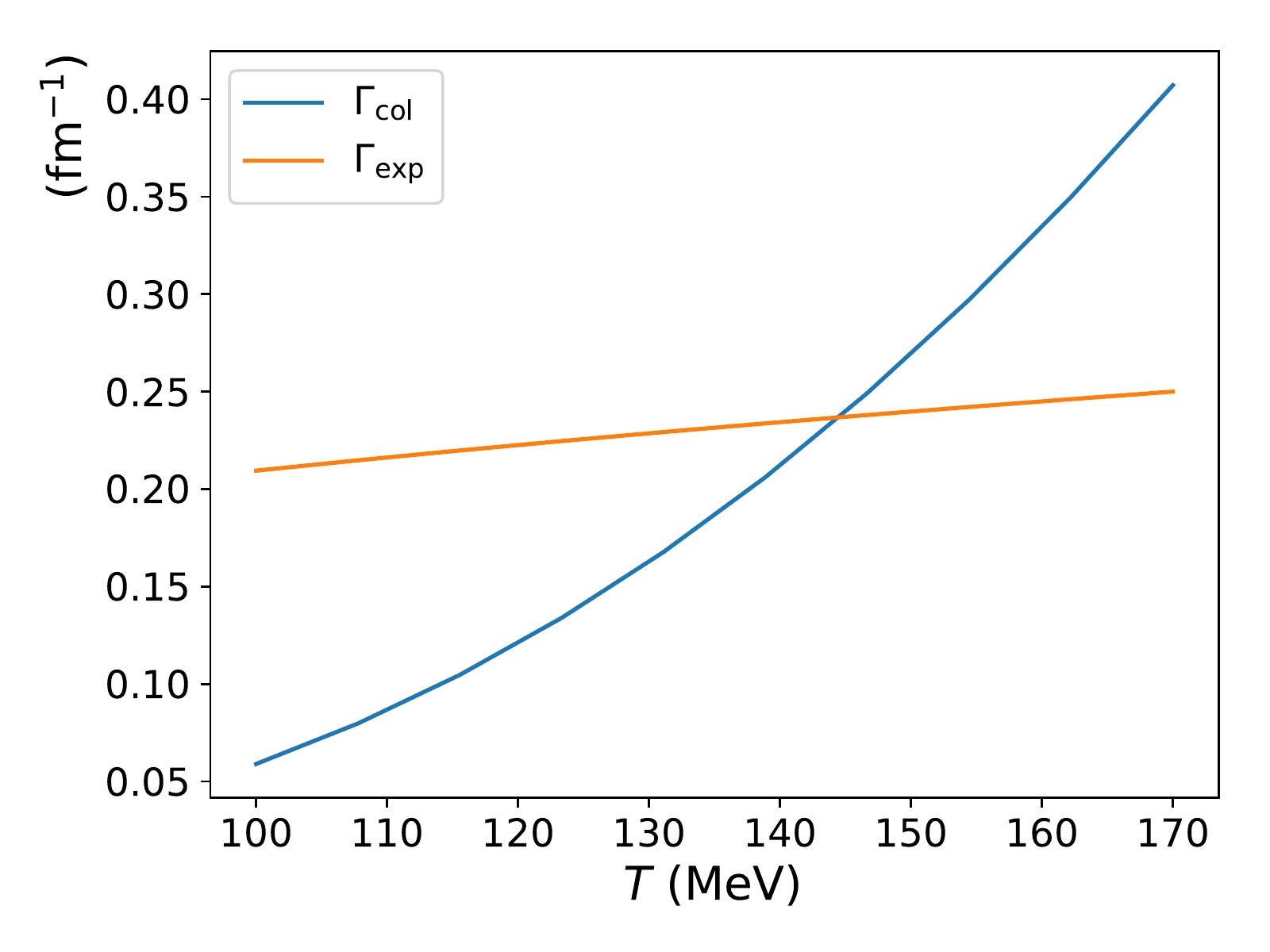}
\caption{\label{fig:rate}  {\bf Left panel:} average mean free path of a pion as function
of the temperature, becoming larger as the system cools and the interactions weaken as dictated by chiral perturbation theory. {\bf Right panel:} The averaged collision rate for an expanding pion gas compared to the Bjorken model's expansion rate, showing that a $T_f=120$ MeV as seen in experiment is a reasonable outcome as the system has become frozen.}
\end{figure}

An external pion probe $i$ of momentum $p_i$ would undergo collisions with the pions (with momentum $p_g$) of an infinite, thermalized medium (of interest for cosmology~\cite{Dobado:2015vaa}, but only a first approximation in heavy-ion collisions), at a rate and with a mean free path given by
\begin{equation}
\label{eq:cr_primer}
    \Gamma (p_i; T)
    =  \int \frac{d^3 p_g}{(2\pi)^3}
      f(p_g; T) \sigma (p_g, p_i) \left| {\bf v}_{\mathrm{rel}} (p_g, p_i) \right|\ ;\ \ \  
\lambda (p_i ; T)   = \frac{p_i}{E_i} \frac{1}{\Gamma (p_i ; T)}\ ,
\end{equation}
where $f(p_g;T)$ is the thermal pion occupation number and ${\bf v}_{\mathrm{rel}}$ the relative velocity of two colliding pions. The average mean free path $\overline{\lambda}(T)$ in such infinite medium is shown in the left panel of Fig.~\ref{fig:rate}.

Because the gas is finite and in expansion, we take this expression as a local approximation only, and convolve it with the expansion of the gas in the Bjorken model. In such expansion the distribution function depends on the rapidity ($Y$) that we will average over in the central region,
\begin{equation}
  f(x,p;T) = \frac{g}{e^{\beta p^{\mu}u_{\mu}}-1 }\ ,\quad
  p^{\mu}u_{\mu}= E \cosh Y - p \cos\theta \sinh Y\ .
\end{equation}
Because the equation of state of the pion gas is approximately linear~\cite{Rapp:1995py} (with squared sound speed $c_s^2 \simeq 0.27 < 1/3$), the temperature as function of the proper (Bjorken) time ($\tau$) is $T(\tau)=T_0 \left( \frac{\tau_0}{\tau}\right)^{c_s^2}$. 
We can fix one of the two constants from data by running the equation backwards: from Hanbury Brown-Twiss interferometry~\cite{ALICE:fo} we know that the freeze-out system size yields about 11 fm $=r_f \simeq \tau_f$ (at zero rapidity), while the pion spectrum can be fit by the kinetic freeze-out temperature $T_f\simeq 120$ MeV. Taking $T_0\simeq 170$ MeV for the transition from the quark-gluon plasma to the hadron gas, the Bjorken evolution is started at $\tau_0\simeq 3$ fm.

The expansion rate in this longitudinal Bjorken expansion model is then 
\begin{equation}
\Gamma_{\rm exp} (T(\tau)) = \partial_\mu u^\mu (\tau) = \frac{1}{\tau_0} \left( \frac{T (\tau)}{T_0} \right)^{1/c_s^2}\ ,
\end{equation}
and kinetic freeze-out (decoupling) of the pion gas happens when this rate is significantly larger $\Gamma_{\rm exp} (T) \gg \Gamma_{\rm col} (T)$ than the pion interaction rate averaged over the gas, defined as
\begin{equation}
   \Gamma_{\rm col}(T) =\frac{1}{n(T)} \int \frac{d^3 p'}{(2\pi)^3}\int \frac{d^3 p}{(2\pi)^3} \sigma(p,p') f(p;T) f(p';T)  |{\bf v}_{\rm rel}(p,p')|\ ,
\end{equation}
where $n(T)$ is the pion density. Both $\Gamma_{\rm exp} (T) $ and $\Gamma_{\rm col}(T)$ are shown in the right panel of Fig.~\ref{fig:rate}.

\begin{figure}[ht]
\includegraphics[width=0.48\columnwidth]{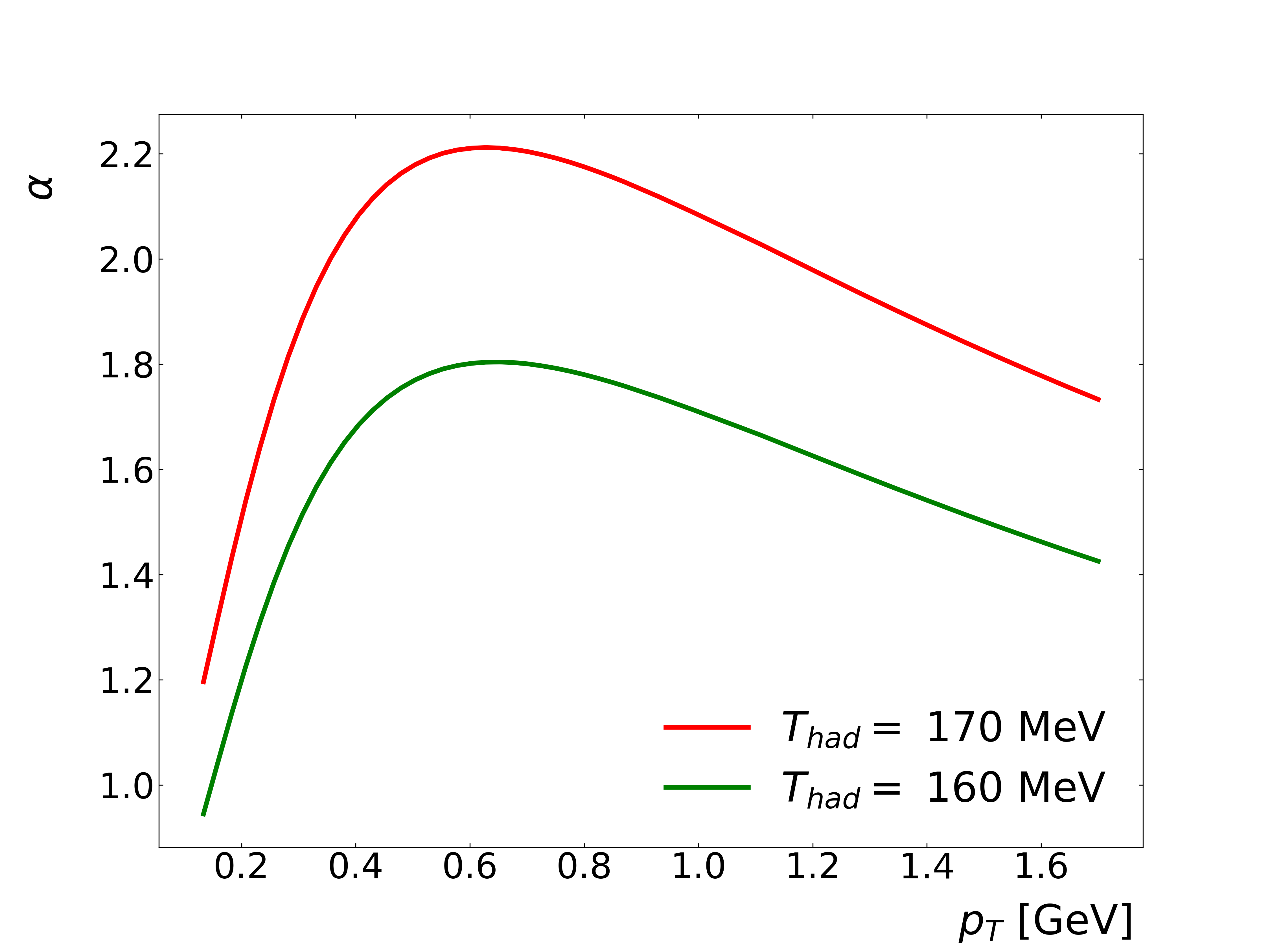}
\includegraphics[width=0.47\columnwidth]{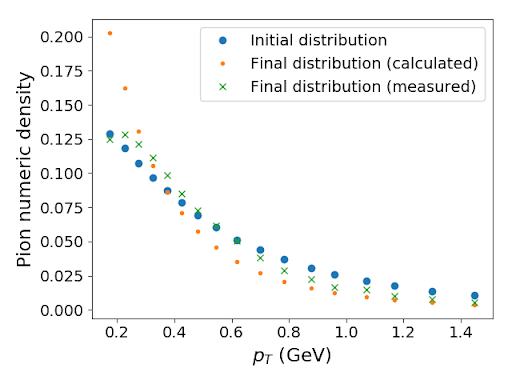}
\caption{\label{fig:pt} 
{\bf Left panel:} Pionic depth of the pion gas in the final stage of (top-energy) RHIC events.
{\bf Right panel:} Distortion of the pion $p_T$ spectrum caused by the terminal phase in (top-energy) RHIC events.}
\end{figure}

The collision rate can alternatively be integrated over the lifetime of the collision yielding the pionic depth (equivalent to the optical depth or inverse opacity in a dense optical medium) as function of pion $p_T$ (at midrapidity),
$\alpha(p_T) \equiv \int_{\tau_1}^{\tau_2} \Gamma(\tau,p_T) d\tau $ (Figure~\ref{fig:pt}, left panel) from which the corresponding fraction of pions that escape without interaction can be extracted, $\mathcal{F} (p_T) = \exp[-\alpha (p_T)]$. 
We plot this fraction  $\mathcal{F}$ in page 1, as it is our highlighted result. It is small, and lowest around $p_T=0.5$ GeV (such pions can collide with the average ones in the medium to form the $\rho$ resonance that disperses them), of order 10-16\% (depending on the value of $T_{had}$ employed). 

In Fig.~\ref{fig:pt} (right panel) we show the distortion of the $p_T$ spectrum of the pion gas brought about by its self-interaction. We observe that the initially-distributed mid-$p_t$ pions ($p_T \sim 1$ GeV) from a thermal-model (blue dots) are pushed to lower $p_T$ (orange dots) that we obtain by subtracting from the initial one a fraction consistent with the left panel, and depositing them again following the final Maxwell-Boltzmann thermal distribution at the freeze-out temperature. The measured distribution (green saltires) does not show such a large proportion of pions around $p_T=0.1$ GeV, but this may be affected by their very low detection and reconstruction efficiency in the ALICE TPC.

In conclusion, we have found the pion gas formed in the final state of heavy-ion collisions to be rather strongly coupled, in agreement with earlier investigations~\cite{Tomasik:2002qt}.

\acknowledgments

This project has received support from the EU’s Horizon 2020 research and innovation programme under grant agreement No 824093;  Deutsche Forschungsgemeinschaft under 411563442 (Hot Heavy Mesons) and 315477589—TRR 211 (Strong-interaction matter under extreme conditions);
MICINN grants PID2019-108655GB-I00, -106080GB-C21 (Spain) and the U. Complutense de Madrid research group 910309 \& IPARCOS.



\begin{thebibliography}{99}
\bibitem{Kaminski:2012bv}
R.~Kaminski {\it et al.}
Nucl. Phys. B Proc. Suppl. \textbf{234} (2013), 253-256;
J.~R.~Pelaez, A.~Rodas and J.~Ruiz De Elvira,
Eur. Phys. J. C \textbf{79} (2019), 1008
doi:10.1140/epjc/s10052-019-7509-6


\bibitem{Dobado:2015vaa}
A.~Dobado, F.~J.~Llanes-Estrada and D.~Rodriguez-Fernandez,
Int. J. Mod. Phys. A \textbf{31} (2016), 1650118
doi:10.1142/S0217751X16501189; 
S.~Sau, S.~Bhattacharya and S.~Sanyal,
Eur. Phys. J. C \textbf{79} (2019), 439
doi:10.1140/epjc/s10052-019-6938-6.

\bibitem{Rapp:1995py}
R.~Rapp and J.~Wambach,
Phys. Rev. C \textbf{53} (1996), 3057-3068
doi:10.1103/PhysRevC.53.3057

\bibitem{ALICE:fo}
    K.~Aamodt \textit{et al.} [ALICE collaboration],
    Phys. Lett. B \textbf{696}, 328-337 (2011)
    doi:10.1016/j.physletb.2010.12.053

\bibitem{ALICE}
    B.~B.~Abelev \textit{et al.} [ALICE collaboration],
    Eur. Phys. J. C \textbf{73}, 2662 (2013)
    doi:10.1140/epjc/s10052-013-2662-9 .


\bibitem{Tomasik:2002qt}
B.~Tomasik and U.~A.~Wiedemann,
Phys. Rev. C \textbf{68} (2003), 034905
doi:10.1103/PhysRevC.68.034905.

\end{thebibliography}
\end{document}